\input epsf
\epsfverbosetrue
\documentclass{epl}

\title{Local spectroscopy of a proximity superconductor at
very low temperature}
\shorttitle{Local spectroscopy of a proximity \ldots}
\author{N. Moussy, H. Courtois \and B. Pannetier}
\institute{Centre de Recherches sur les Tr\`es Basses 
Temp\'eratures - C.N.R.S. in convention with Universit\'e Joseph 
Fourier, 25 Avenue des Martyrs, 38042 Grenoble Cedex, 
France}

\begin{document}

\maketitle 

\begin{abstract}
We performed the local spectroscopy of a Normal-metal--Superconductor (N-S)
junction with the help of a very low temperature (60 mK) Scanning Tunneling
Microscope (STM).  The spatial dependence of the local density of states
was probed locally in the vicinity of the N-S interface.  We observed
spectra with a fully-developed gap in the regions where a thin normal metal
layer caps the superconductor dot.  Close to the S metal edge, a clear
pseudo-gap shows up, which is characteristic of the superconducting
proximity effect in the case of a long normal metal.  The experimental
results are compared to the predictions of the quasiclassical theory.
\end{abstract}

In the recent years, there have been a noticeable renewed interest in the
proximity effect appearing at the junction between a Normal metal (N) and a
Superconductor (S).  In particular, the transport properties of the normal
metal part were shown to exhibit a striking energy dependence with a long
range persistence \cite{Revue_JLTP}.  Another aspect of the proximity
effect is the local modification of the energy spectrum.  In the case of a
N-S junction with a finite normal metal, the density of states in the
normal metal is expected to exhibit a mini-gap : the density of states is
zero within an energy window around the Fermi level \cite{Golubov,Belzig}. 
The width of this mini-gap is smaller than the intrinsic superconducting
gap of the superconductor.  If two superconductors are connected to a short
normal metal (S-N-S junction), the mini-gap will moreover depend
on the phase difference between the two superconductors \cite{Zhou}.  In
the case of a normal metal with an infinite length, the density of states
should exhibit a pseudo-gap \cite{Golubov,Belzig} : the density of states
is zero only at the exact Fermi level and goes approximately linearly with
the energy close to the Fermi level.  This behavior can be understood by
arguing that some electron trajectories that travel close the interface
never hit it, therefore do not couple to superconductivity and contribute
to the density of states.  Practically, the criteria of an infinite length
for the normal metal should be understood as a length larger than the phase
coherence length.

The local density of states in the vicinity of a N-S junction has been
measured with the help of submicron planar tunnel junctions \cite{Gueron}. 
The appearance of a pseudo-gap in the N metal as well as the inverse
proximity effect on the S metal side \cite{Sillanpa} were observed.  A more
local study is welcome in order to overcome the unavoidable spatial
averaging of such conventional experiments.  Indeed, the Scanning Tunneling
Microscopy (STM) in the spectroscopy mode enables one to measure the very
local density of states under the tip.  Due to the technical complexity of
(very-)low-temperature STM, the local study of mesoscopic superconductors
is still in its infancy.  Inoue and Takayanagi measured the tunneling
spectra of a Nb-InAs-Nb system at 4.2 K \cite{Takayanagi}.  Pioneering work
by Tessmer et al.  focused on the proximity effect in Au nano-sized islands
on top of a NbSe$_{2}$ sample \cite{vanHarlingen}.  Levi et al.  studied
complex Ni-Cu-NbTi multifilamentary superconducting wires \cite{Millo}. 
Recently Vinet et al.  performed low-temperature spatially-resolved
spectroscopy of Nb-Au structures patterned by lithography \cite{Vinet}.  In
this letter, we report on local spectroscopy measurements we performed on a
proximity superconductor.  We have been able to follow the evolution of the
density of states from a fully-developed gap to a pseudo-gap as we move
from the superconductor to the normal metal.  Close to the N-S interface, a
clear proximity effect is observed, both on the normal metal side and on
the superconductor side of the interface.  Eventually, we compare our
experimental results to the predictions of the quasiclassical theory based
on the Usadel equations.

Compared to previous works, we have been able to combine the very-low
temperature (T $<$ 1 K) conditions with the local probe technique.  This
guarantees a much improved energy resolution, and therefore the possibility to
probe the proximity effect on a larger length scale, together with a high
spatial resolution.  Our very low temperature STM works at 60 mK in a
dilution refrigerator \cite{RSI_STM}.  It features both an atomic
resolution and a large scanning range of $6 \times 6 \mu$m$^2$ at low
temperature.  In the spectroscopy mode, this STM has shown an unprecedented
energy resolution of 36 $\mu$eV. This corresponds to an effective
temperature of 210 mK which had to be introduced in the BCS fit of the
spectroscopy data for plain Al and Nb layers.  These fits were performed
without any inelastic scattering parameter \cite{Dynes}.

Our samples were made by successive in-situ evaporation of Nb
(Superconducting below about 9 K) and Au or Cu (Normal metal).  First, the
Si sample substrate was introduced in the UHV evaporator with a patterned Si
membrane clamped on it.  This 5 $\mu$m-thick Si membrane was previously
patterned by e-beam lithography and deep Reactive Ion Etching.  An array of
circular holes with a periodicity of 4 $\mu$m and a diameter of about 
1.5 $\mu$m were drilled throughout the membrane.  During Nb evaporation, the Si
membrane acted as a mechanical mask, so that only dots of Nb are deposited
on the substrate.  After deposition of 40 nm of Nb, the mask was removed in
situ with the help of a UHV manipulator.  Afterwards, an uniform layer of
20 nm of Cu or Au was deposited.  The pressure was below $10^{-8}$ mbar
during the few minutes between the two evaporations.  This provides a highly
transparent Nb-Au interface.  Let us point out that no lithography or
two-step deposition procedure was needed here, so that the Nb-Au interface
is actually as clean as possible.  We will concentrate here on one of the
two Nb-Au samples we studied.  Au was preferred in order to ensure a
natural low oxidation of the sample surface during the transfer from the
evaporator to the cryostat.  One Nb-Cu sample showed a behavior similar to
the Nb-Au samples.  However, the more oxidized surface of Cu rendered the
tunnel junction less stable and therefore more noisy.

\begin{figure}
\onefigure{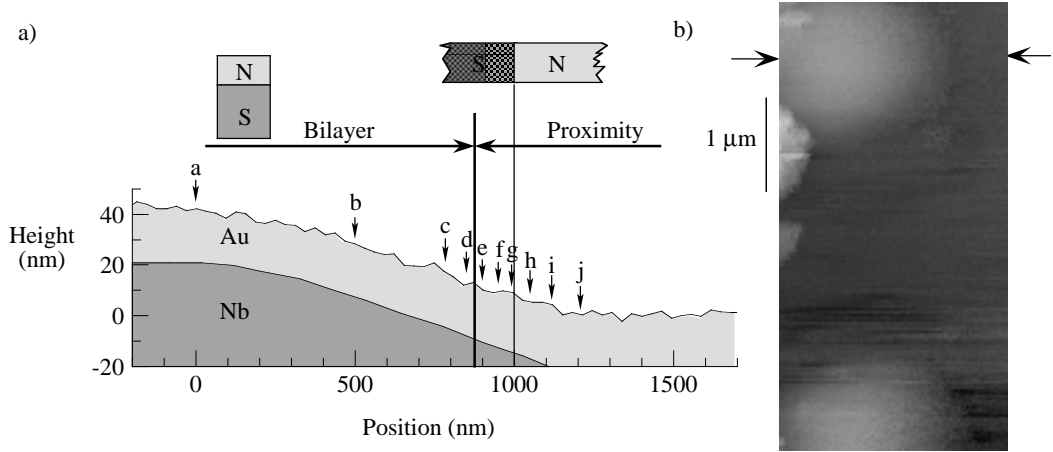}
\caption{a : Profile of the sample surface extracted from image b) (line
indicated indicated by the two arrows) together with a schematics of the
sample geometry.  Note that the vertical scale has been expanded by a
factor of about 10 compared to the horizontal one.  The locations where the
spectra of Fig.  \ref{Spectres}a were measured are indicated, as well as
the domains of application of the two geometry models used in the
calculations (shown in Fig.  \ref{Spectres}b).  In the "Bilayer" region,
the sample is modeled by a series of (vertical) N-S bilayers, the thickness
of S = Nb being given by the measured profile.  In the "Proximity" region,
the sample is modeled by a single (horizontal) N-S region.  b : $2.6 \times
5 \, \mu$m$^2$ STM image at 60 mK captured with a 10 mV bias voltage and a
30 pA tunnel current.  Two circular Nb dots are visible.}
\label{Geometry}
\end{figure}

\begin{figure}
\twoimages[scale=1]{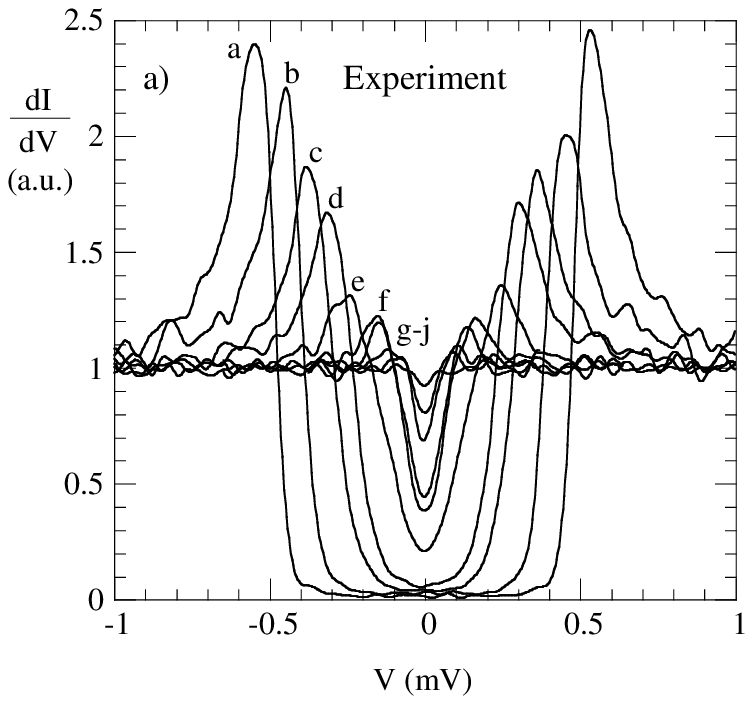}{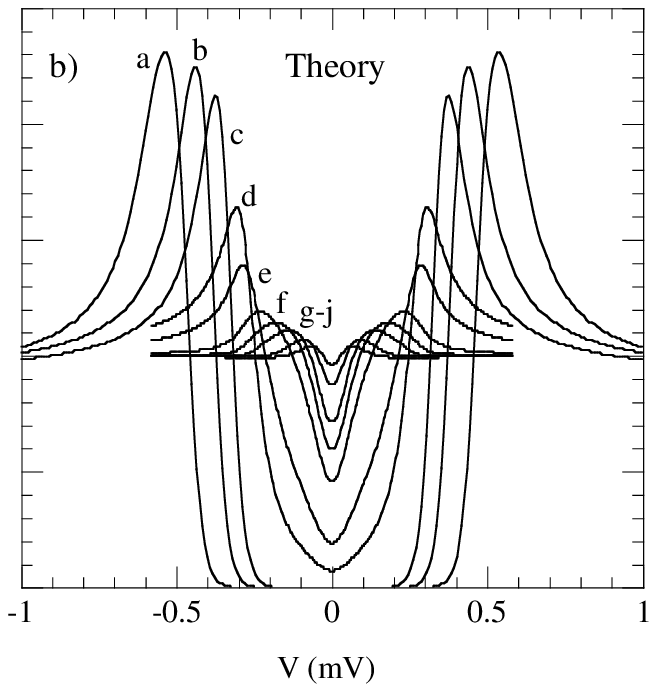}
\caption{a : Experimental spectra measured on locations a to j (see Fig. 
\ref{Geometry}a) during a single scan along one line.  The tunnel
resistance was about 12 M$\Omega$ during the spectroscopies.  b :
Theoretical spectra calculated with the Usadel equations.  The actual
geometry was modelled as a series of N-S bilayers (curves a to d) and a
single lateral N-S junction (curves e to j).  An effective temperature of
210 mK was introduced.  In the bilayer region, the measured thickness 
of the Nb layer ($d_{S}$ = 40, 21, 15 and 11 nm in a, b, c and d respectively) 
in units of $\delta_{S}$ = 27 nm was introduced in the calculation 
together with the fixed Au layer thickness $d_{N}$ = 20 nm = $0.3 \delta_{N}$. In 
the proximity-effect region, the distance from
the interface was a free fit parameter (see Fig. \ref{Longueurs}b).}
\label{Spectres}
\end{figure}

Fig.  \ref{Geometry}b shows an STM image of the Nb-Au sample at 60 mK. Two
Nb dots are clearly visible.  The relief of the Nb dots is rather smooth. 
This is due to both the thickness of the mask and the residual distance
between the substrate and the mask.  This smooth relief is a drawback in
terms of geometry simplicity but an advantage for STM imaging.  We
performed several series of spectroscopies as a function of distance from
the center of a Nb dot by traveling along one line.  During each series,
the displacement speed was reduced to 10 nm/s and the scanning direction
was kept fixed in order to reduce the piezo-electric hysteresis.  Fig. 
\ref{Spectres}a shows a representative selection of spectra taken during a
single series.  The surface profile of the same line is shown in Fig. 
\ref{Geometry}b.  Labeled arrows indicate the position of the spectra shown
in Fig.  \ref{Spectres}.  In the center of the Nb island (curve a), the
density of states exhibits a clear gap, which is reminiscent of a BCS
behavior.  Compared to the bulk Nb gap value $\Delta_{Nb} \simeq 1.4$ meV,
the measured gap is significantly reduced.  This behavior is consistent
with the measured critical temperature $T_{c}$ = 3 K at the
superconductivity onset.  As we move away from the Nb dot center (curves b
to d), the density of states first continues to exhibit a fully developed
but reduced gap.  This remains approximately true up to close to the curve
e, which shows a clear pseudo-gap : the density of states goes
approximately linearly to zero at the Fermi level.  As the tip is moved
further away (spectra e to j), the pseudo-gap width is reduced.  We have
been able to observe a pseudo-gap behavior in the density of states spectra
over about 300 nm.

The proximity effect in diffusive metals can be described by the
quasiclassical theory based on the Usadel equations
\cite{Golubov,Belzig,Zhou,Bruder,Feigelman}.  In the usual $\theta$
parametrization, the complex angle $\theta(\epsilon,x)$ is related to the
pair amplitude as : $F(\epsilon,x)=-i\sin\theta(\epsilon,x)$.  The local
density of states is expressed as :
\begin{equation}
    n(r,\epsilon) = n_{0}\Re e[cos\theta(r,\epsilon)].
\end{equation}
  The Usadel
equations write :
\begin{equation}  
    \left\{
\begin{array}{ll}    
   \frac{1}{2}\hbar D_{S}\delta_{r}^2\theta + i\varepsilon \sin \theta + 
   \Delta(r) \cos\theta = 0
& in \, S\\
   \\
   \frac{1}{2}\hbar D_{N}\delta_{r}^2\theta + i\varepsilon \sin \theta = 0
   & in \, N
\end{array}
\right.
\end{equation} 
where $D_{N}$ and $D_{S}$ are the diffusion coefficients in N and S
respectively.  The inelastic and spin-flip rates were neglected.  The gap
$\Delta(r)$ in S is self consistently defined by :
\begin{equation}
    \Delta (r) = n_{0}V_{eff}\int_{0}^{\hbar w_{D}} 
    tanh(\frac{\varepsilon}{2 k_{B}T}) \Im m [\sin \theta] d\varepsilon,
\end{equation}
where $n_{0}$ is the electron density, $V_{eff}$ is the local interaction
parameter and $\omega_{D}$ is the Debye frequency.  In the case of a
perfect transparency, the boundary conditions at the N-S interface include
the continuity of $\theta$ at the interface and the spectral current
conservation : $\sigma_{S} (\partial \theta / \partial
x)_{x=0^{-}}=\sigma_{N}(\partial \theta / \partial x)_{x=0^{+}}$, where
$\sigma_{S}$ and $\sigma_{N}$ are the conductivity in S and N respectively. 
In this work, we benefited from the numerical code developed by W. Belzig
{\it et al.} \cite{Belzig}, which solves the Usadel equation for a quasi-1D
N-S junction and calculates the local density of states.  The relevant
theoretical parameters are : the gap $\Delta$ of the S metal, the mismatch
parameter $\gamma = \frac{\sigma_{N}}{\sigma_{S}}
\sqrt{\frac{D_{S}}{D_{N}}}$ and the thicknesses of the N and S layers in
units of the characteristic lengths $\delta_{N}=\sqrt{\hbar D_{N}/2\Delta}$
and $\delta_{S}=\sqrt{\hbar D_{S}/2\Delta}$ respectively.  Whereas the
length $\delta_{S}$ is close to the superconducting coherence length
$\xi_{S}$, the length $\delta_{N}$ does not stand for a coherence length of
electron pairs in the normal metal.  In order to analyze easily our
experimental results, we modeled the different parts of our structures as
quasi-1D junctions.  On the Nb dot, we locally model the sample as an
uniform bilayer of Nb and Au, see Figure \ref{Geometry}b.  Note that the
vertical magnification by a factor of about 10 may make the reader
underestimate the validity of this approach.  In the remaining region
labeled "Proximity", we model our sample as a single horizontal N-S
junction, the interface position being {\it a posteriori} determined by the
comparison with the theory.

\begin{figure}
\twoimages[scale=1]{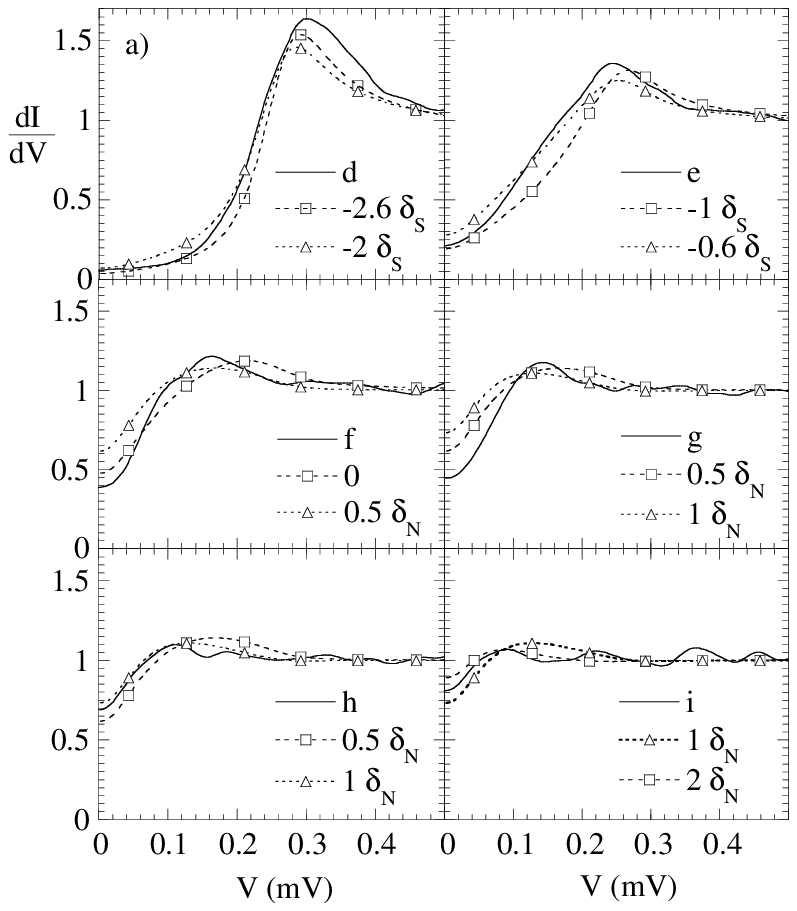}{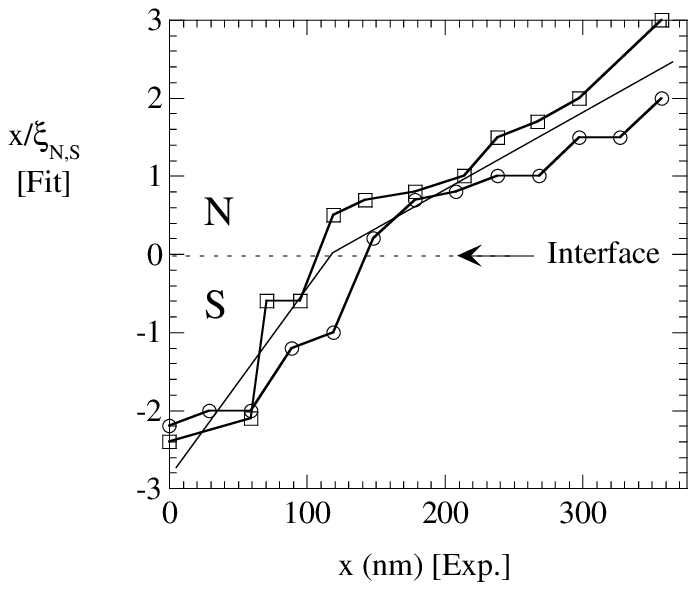}
\caption{a : Comparison between individual experimental data and two
related calculated curves for an N-S junction with infinite N and S metals,
using the parameters $\gamma=1.1$ and $\Delta$ = 0.27 meV. The position
compared to the N-S interface is expressed in units of the characteristic
length scales $\delta_{N}$ and $\delta_{S}$.  b : Comparison of the
fit-derived distance from the interface with the actual (measured)
position.  Two data series taken on two different lines are shown and
identified by squares and triangles symbols.  The slopes of the adjusted
lines give the length scales values $\delta_{S}$ = 50 nm and
$\delta_{N}$ = 94 nm.}
\label{Longueurs}
\end{figure}

Let us first consider the "Bilayer" region.  The sample is locally modeled
as a quasi-1D N-S junction with a constant length of N metal (the 20 nm
thickness of the N layer) and a locally varying length of S metal (measured
in the experiment).  Since the Normal metal is small, a fully-developed gap
is predicted, in agreement with the experiment.  We assumed a perfect
interface transparency and no inelastic or spin-flip scattering.  By
fitting as a first step only the curve a, a good set of values was found to
be $\gamma=1.1$, $\Delta$ = 0.82 meV, $d_{S} = 1.5 \delta_{S}$ and 
$d_{N} = 0.3 \delta_{N}$.  The value of the $\gamma$ parameter matches within the
experimental accuracy the estimation of 1.2 based on the measured transport
properties of the N and S layers.  The magnitude of $\Delta$ is
significantly smaller than the Nb bulk value, which we attribute to the
small Nb thickness \cite{Geballe} and the special conditions of Nb
evaporation.  As the Nb and Au layers thicknesses are known, we can extract
the length scales $\delta_{S}$ = 27 nm and $\delta_{N}$ = 67 nm.  These
values are in qualitative agreement with the respective expected values 
41 nm and 55 nm.  In this respect, let us note that the assumption of a
diffusive motion over each layer thickness is only partially fulfilled
since the mean free paths are of the order of the layer thicknesses. 
Curves for locations b to d in the "Bilayer" region were subsequently
calculated by taking into account the fixed Au thickness $d_{N}$ = 20 nm
= $0.3 \delta_{N}$ and the measured Nb layer thickness in units of
$\delta_{S}$ = 27 nm (inferred from curve a fit).  As expected, the
mini-gap amplitude is reduced as the thickness of the Nb layer decreases,
the Au layer thickness being constant.  The calculated curves are shown in
Fig.  \ref{Spectres}b.  In every curve, we introduced an effective
temperature T = 210 mK in order to account for the experimental accuracy
\cite{RSI_STM}.  The agreement is very good, as both the mini-gap amplitude
and the overall spectra shape are well described.

In the "Proximity" region, the measured spectra (curve e and beyond) show a
filling of the density of states near the Fermi level and a peak shape that
are not compatible with a bilayer model.  This pseudo-gap behavior is the
signature of the proximity effect in a N-S junction with a long normal
metal N. In this "Proximity" region, we described the sample as a single
infinite N-S junction extending laterally.  We used the same $\gamma$
parameter as in the "Bilayer" region.  Again we consider a perfect
interface transparency and zero inelastic and spin-flip scattering.  In
order to describe successfully the data, we had to assume a reduced gap
value $\Delta'$ = 0.27 meV. Compared to the value in the bilayer region,
this reduced value may be understood as the effective gap of the Nb-Au dot
treated as a whole.  The experimental curves e to j were fitted within this
model by considering the tip position as a free parameter.  As an
illustration, Fig.  \ref{Longueurs} shows the comparison between one
experimental data and two theoretical curves which differ in the position
compared to the interface. This position is expressed in units of the
characteristic length scales $\delta_{N,S}$. The main fitting criteria was
the density of states peak position, which actually moves towards the Fermi
level as the position $x$ referred from the interface is increased.

The validity of our description can be checked in Fig.  \ref{Longueurs}b
where we compare the actual position of the acquired spectra to the
position extracted from the fit, in units of the relevant characteristic
length.  The data points follow a monotonous behavior, but with a
significant scattering.  The change of slope at the N-S interface reflects
the difference in the characteristic length scales $\delta_{N}$ and
$\delta_{S}$.  From the slope of the mean line we can draw through the data
points on the N side (top part of Fig.  \ref{Longueurs}), we can extract
the value $\delta_{N}$ = 94 nm.  This corresponds exactly to the estimation
based on the gap $\Delta'$ and the measured mean free path of 16 nm in Au. 
On the S side (bottom part of Fig.  \ref{Longueurs}), the estimated length
is $\delta_{S}$ = 50 nm.  Taking into account the
reduced gap $\Delta'$, it corresponds to a mean free path of 4.5 nm which
is half the value estimated from transport properties of similar samples. 
In fact, it should be considered more as a property
of the Nb-Au layer at its border than a property of the bare Nb film.  This
means that although our description seems both accurate and
well-established in the N region, it is more difficult to relate the
theoretical description on the S side to the real physical situation.

The location of the interface can be accessed from the fitting procedure. 
It appears that the interface is situated at an estimated Nb thickness of 6
nm.  This means that the Nb layer is not superconducting when it is thinner
than 6 nm.  This behavior is expected in respect of the getter properties
of Nb during evaporation.  Let us also note that the density of states
close to the interface is already much depressed compared to the "bulk"
density of states in the superconductor.  From the theory, this behavior is
very sensitive to the mismatch parameter $\gamma$.  This parameter is not
intrinsic of the two materials since it not only depends on the N and S
electronic densities ratio but also the mean free paths ratio.  In Fig. 
\ref{Longueurs}, the scatter in the fitted position compared to the actual
position can be envisioned as a signature of the extreme locality of the
STM measurement.  It is indeed visible from our whole set of data 
that the measured spectrum often changed abruptly as the surface
is scanned.  Some of these events are really local at the nanometer scale
and show anomalies presumably related to point-like impurities.  The other
events feature step-like evolutions which are indeed in agreement with the
overall behavior expected from the proximity effect.  This behavior may be
related to local scattering centers in the sample, like grain boundaries.

In conclusion, we have been able to probe the local density of states in
the vicinity of a N-S junction at very low temperature.  On the
superconductor side of the junction, we observed a strong inverse proximity
effect, with a density of states strongly modified compared to the BCS
spectra.  On the normal metal side, we observed a pseudo-gap of decreasing
amplitude as the tip moves away from the N-S interface.  The experimental
data compares favorably with the spectra calculated from the Usadel
equations, assuming a very simple geometrical model of the complex sample
geometry.  In contrast with previous studies \cite{Gueron, Vinet}, we found
that spin-flip and inelastic scattering could be neglected.  This means
that the related characteristic lengths should be larger than about 300 nm.

\acknowledgments
We gratefully thank W. Belzig for sharing his numerical code for the
resolution of the Usadel equation with us.  We acknowledge the help of T.
Fournier and T. Crozes for the Si membrane technique and discussions with
V. Chandrasekhar and L. Cr\'etinon.  We acknowledge discussions within the
TMR network "Dynamics of Superconducting Nanocircuits" and the "Mesoscopic
Electronics" COST network.  This work is supported by the French Ministry
of Education and Research under an "ACI Nanostructures" contract.

\end{document}